# Optical rogue wave in random distributed feedback fiber laser

Jiangming Xu[1,2], Jian Wu[1,2], Jun Ye[1], Pu Zhou[1,2,*], Hanwei Zhang[1,2], Jiaxin Song[1], & Jinyong Leng[1,2]

The famous demonstration of optical rogue wave (RW)-rarely and unexpectedly event with extremely high intensity-had opened a flourishing time for temporal statistic investigation as a powerful tool to reveal the fundamental physics in different laser scenarios. However, up to now, optical RW behavior with temporally localized structure has yet not been presented in random fiber laser (RFL) characterized with mirrorless open cavity, whose feedback arises from distinctive distributed multiple scattering. Here, thanks to the participation of sustained and crucial stimulated Brillouin scattering (SBS) process, experimental explorations of optical RW are done in the highly-skewed transient intensity of an incoherently-pumped standard-telecom-fiber-constructed RFL. Furthermore, threshold-like beating peak behavior can also been resolved in the radiofrequency spectroscopy. Bringing the concept of optical RW to RFL domain without fixed cavity may greatly extend our comprehension of the rich and complex kinetics such as photon propagation and localization in disordered amplifying media with multiple scattering.

[1] College of Advanced Interdisciplinary Studies, National University of Defense Technology, Changsha 410073, China . [2] Hunan Provincial Collaborative Innovation Center of High Power Fiber Laser, Changsha 410073, China. J. X. and J. W. contributed equally to this paper. Correspondence and requests for materials should be addressed to P. Z. (email: zhoupu203@163.com).

The concept of random laser employing multiple scattering of photons in an amplifying disordered material to achieve coherent light has attracted increasing attention due to its special features, such as cavity-free, structural simplicity, and promising applications in imaging, medical diagnostics, and other scientific or industrial fields [1-3]. Meanwhile, because of the lack of sharp resonances, it also has presented many challenges to conventional laser theory, such as light localization phenomenon in the disordered amplifying media [4-6]. Moreover, random fiber laser (RFL), whose operation is based on the extremely weak Rayleigh scattering (RS) provided random distributed feedback in a piece of passive fiber, can trap the random laser radiation in one-dimensional waveguide structure with efficiency and performance comparable to conventional fiber lasers [7, 8]. It also presents rich physical properties in spectral, temporal and spatial domains [9-11], especially the intensity statistical characteristics as an important gate for understanding the operational mechanism and defining features of light source [12-15].

In general, intensity statistical investigations of fiber sources have experienced a flourishing time since the famous demonstration of optical rogue wave (RW) [16]. Although the statistical RW behavior with (temporally or spatially) highly-skewed intensity distributions and rarely and unexpectedly appearance has been known for a long time in various different physical contexts as oceanographic wave, capillary waves, plasma waves and Bose–Einsteins condensates (see, e.g., [17, 18], for extensive reviews of this field), the optical RW phenomenon was first presented, yet, in 2007 by Solli et al. when they investigated the heavy-tailed histograms of intensity fluctuations in supercontinuum generation [16]. In recent years, the mechanisms of formation of optical RWs have been done both experimentally and theoretically in fiber lasers with nonlinearly driven cavities [16, 19], Raman fiber amplifiers and lasers [20, 21], and fiber lasers via modulation of the pump [22] (see the comprehensive overviews on this issue [23, 24]). In contrast, up to now, the optical RW behavior with temporally localized structure has not been presented in RFL with no defined cavity (or with an open cavity), even if stochastic pulse shape origin from stimulated Brillouin scattering (SBS) has been observed in RFL [7, 25-27]. Detailed, in the former RFLs constructed by a piece of common passive fiber, the SBS factor and associated giant pulses can only exist near the lasing threshold and will be suppressed in the power scalability (typically, about 25 % higher than lasing threshold in [7]) by the potential intensity fluctuations of pump laser, as discussed in [26, 28]; otherwise, a section of special ultra-high numerical aperture (UHNA) passive fiber with small core diameter (usually about 2-4 μm), limited conversion efficiency (about 10%) and robustness, is necessary for the SBS stimulation [25, 27]. Consequently, the generation and evolution of optical RW behavior in RFL has until now been an open issue.

Here, we demonstrate for the first time to our knowledge the optical RW behavior in RFL sustained from nearly lasing threshold to even maximal pump power (about 2.5 times higher than the lasing threshold) by utilizing a simple experimental setup along with a piece of standard telecom fiber span, in combination with an incoherently amplified spontaneous emission (ASE) source with high temporal stability as pump source. Furthermore, threshold-like beating-peak phenomenon in the RF spectroscopy is also shown and discussed initially. The observation of temporally localized events in this cavity-free structured RFL may pave a new way to reveal the complex fundamental physics underlying the lasing from multiple scattering in one-dimensional disordered amplifying media.

**Results**

**Experimental strategy and primary output characteristics.** The proposed principle of the incoherently pumped self-pulsed RFL operation is depicted in Fig. 1. It's well known that incident photons propagating in the long passive fiber can be backscattered by the random refractive index inhomogeneity induced weak RS and amplified by the stimulated Raman scattering (SRS) in this cavity-free RFL. Simultaneously, the acoustic field generated as a result of electrostriction can induce moving and stochastic density grating, which is defined as SBS process [29]. Then, frequency downshifted photon owing to Doppler Effect will be stimulated with temporal transient characteristic. Consequently, self-pulsed random laser can be achieved with the participation of SBS effect.

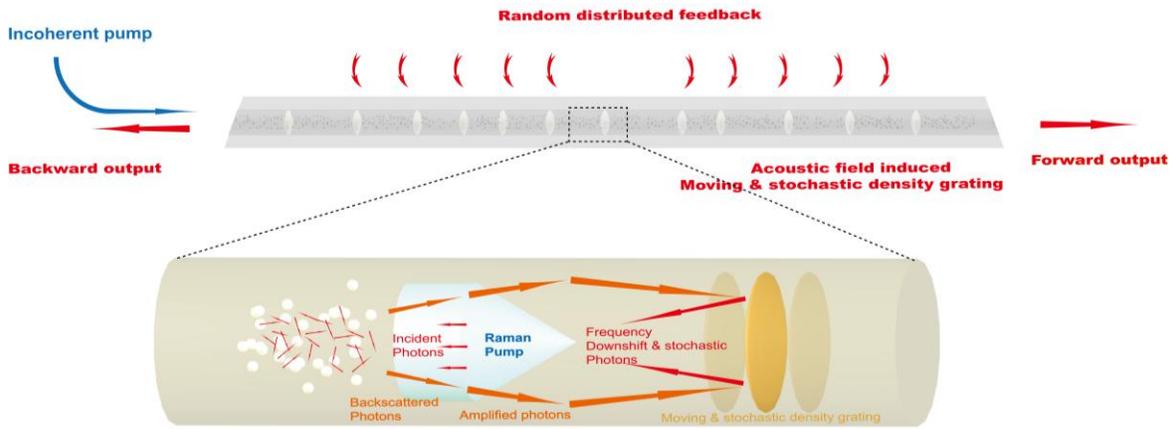

Figure 1 | Operating principle of the incoherently pumped random fiber laser (not to scale).

Figure 2 illustrates the evolutions of output power and spectrum with the boosting of pump level, which can clearly demonstrate the threshold behavior. Below the random lasing threshold (about 6.11 W), most of the pump light is untransformed and transmitted to the forward direction, and the power of spontaneous Raman emission can be neglected, as plotted in Fig. 2(a). The spontaneous Raman emission spectrum is presented in Fig. 2(b), corresponding to a full width at half maximum (FWHM) linewidth of about 12.54 nm. With the enhancing of pump power over threshold, random intense spikes localized around the Raman gain profile maxima (1120 nm) can be observed, which has a contrast of ~30 dB against the spontaneous Raman emission pedestal. Simultaneously, due to the typically Schawlow-Townes spectral narrowing effect of random distributed feedback fiber laser near the lasing threshold [9, 30], the FWHM linewidth of output spectrum dramatically decreases to an average value of 0.03 nm. Higher pumping results in the nonlinear increasing of output random laser power. With ultimate 21.29 W pump power employed (limited by the pump source), maximal output powers of 8.14 W and 3.44 W can be obtained for the backward and forward $1^{st}$ order Stokes light, respectively. The backward $1^{st}$ order Stokes light power is significantly higher than that of forward output, which can be explained as following. Firstly, due to the higher pump power distribution near the backward output port, the Raman gain for the Stokes light are higher. Secondly, despite the Stokes light can be generated and amplified in both forward and backward directions, the forward Brillouin scattering in optical fiber is very weak, and in hence the energy conversion from the pump light to the backscattered wave is more efficient [29]. As to the output spectrum, broadband spectral envelope can be observed with mass characteristic spectral spikes even under the maximal operation power, and the separations between spectral spikes are ~0.06 nm as expected for Stokes shifted SBS in 1 μm range [19]. Besides, anti-Stokes components shorter than 1120 nm are also generated through four-wave-mixing effect [31]. Furthermore, the spectral envelope varies from time to time with random spectral peaks and envelope linewidth, which can be attributed to the typical stochastic spectral behavior of SBS effect [32] as the composed spectral peaks mainly originate from SBS scattering. To characterize the fluctuation of spectral envelope quantitatively, we depict the minimum, maximum and mean values of 20 measurements in Fig. 2(c). Generally speaking, obviously spectral envelope broadening can be observed in the power scaling process for the nonlinear effects, such as high-order SBS generation, self-phase modulation (SPM) and cross-phase modulation (XPM) [9]. Additionally, the statistical mean value of FWHM linewidth evolves to a stable value of about 14 nm with the enhancement of pump power over 12.65 W.

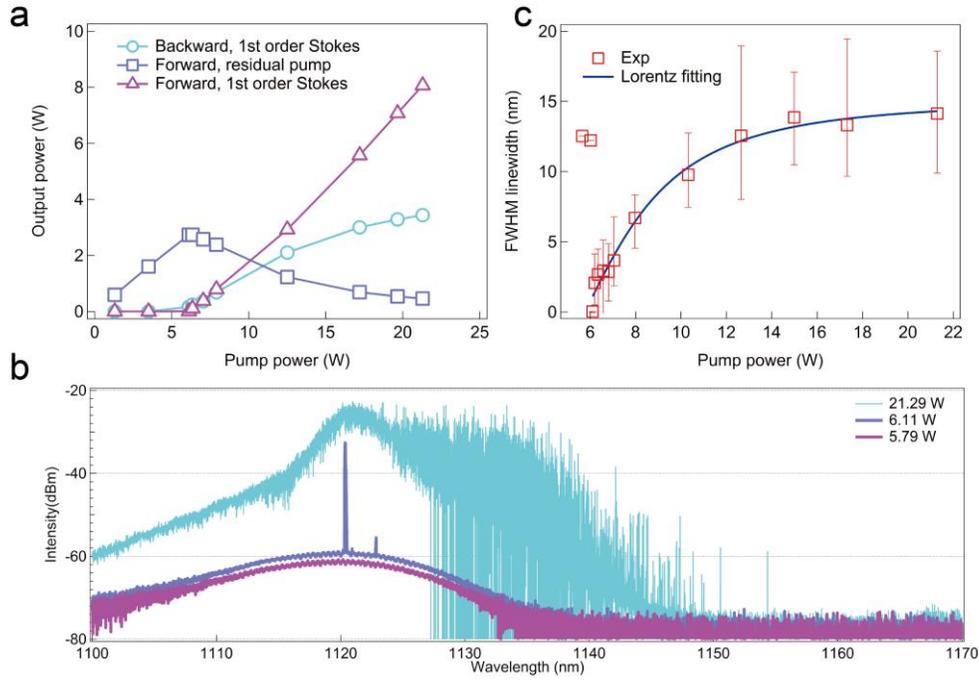

Figure 2 | Evolutions of output power and spectrum. (a) The output powers of 1st order Stokes light and residual pump light as functions of pump power. (b) Typical spectra of the backward 1st order Stokes light at different pump levels. The spectra are recorded with a resolution of 0.02 nm. (c) FWHM linewidth of backward spectral envelope as a function of injected pump power. The circles indicate the average FWHM linewidth of 20 measurements. The corresponding error bars remark the variation range of FWHM linewidth at fixed pump level.

**Observation of rogue wave behavior.** As we know, pulsation operation can be obtained as the SBS factor can switch the Q value of RFL [25, 27, 33] and exist in the power boosting process. The evidence and evolution of SBS effect will be given in the following with the aid of radio frequency measurement at corresponding operation powers. To analyze the temporal characteristics of output pulsed random laser, we employ a high-speed photodetector (45 GHz bandwidth) and a 16 GHz real-time oscilloscope with 100 MPts/channel capacity allowing to record time series up to 32 ms with a 320 ps/point resolution in order to have sufficient statistics. The typical temporal trace of output random laser around threshold is recorded and plotted in Fig. 3(a), in which a large set of pulse clusters could be observed. The closed-up view of pulse bundle with the highest amplitude is plotted in Fig. 3(b). Induced by the intrinsic stochastic nature of SBS effect, unstable pulsations are observed with random intervals, durations and amplitudes. Furthermore, most of the pulses are so weak that they are nearly buried beneath the noise floor; however, the most giant ones can reach intensities about 20 times the lowest. With the scaling of pump power from lasing threshold to maximal level, the pulse peak amplitude increases, and the giant pulses become denser (the number of pulses for a fixed time interval increases), as shown in Fig. S1 of the supplementary Figures.

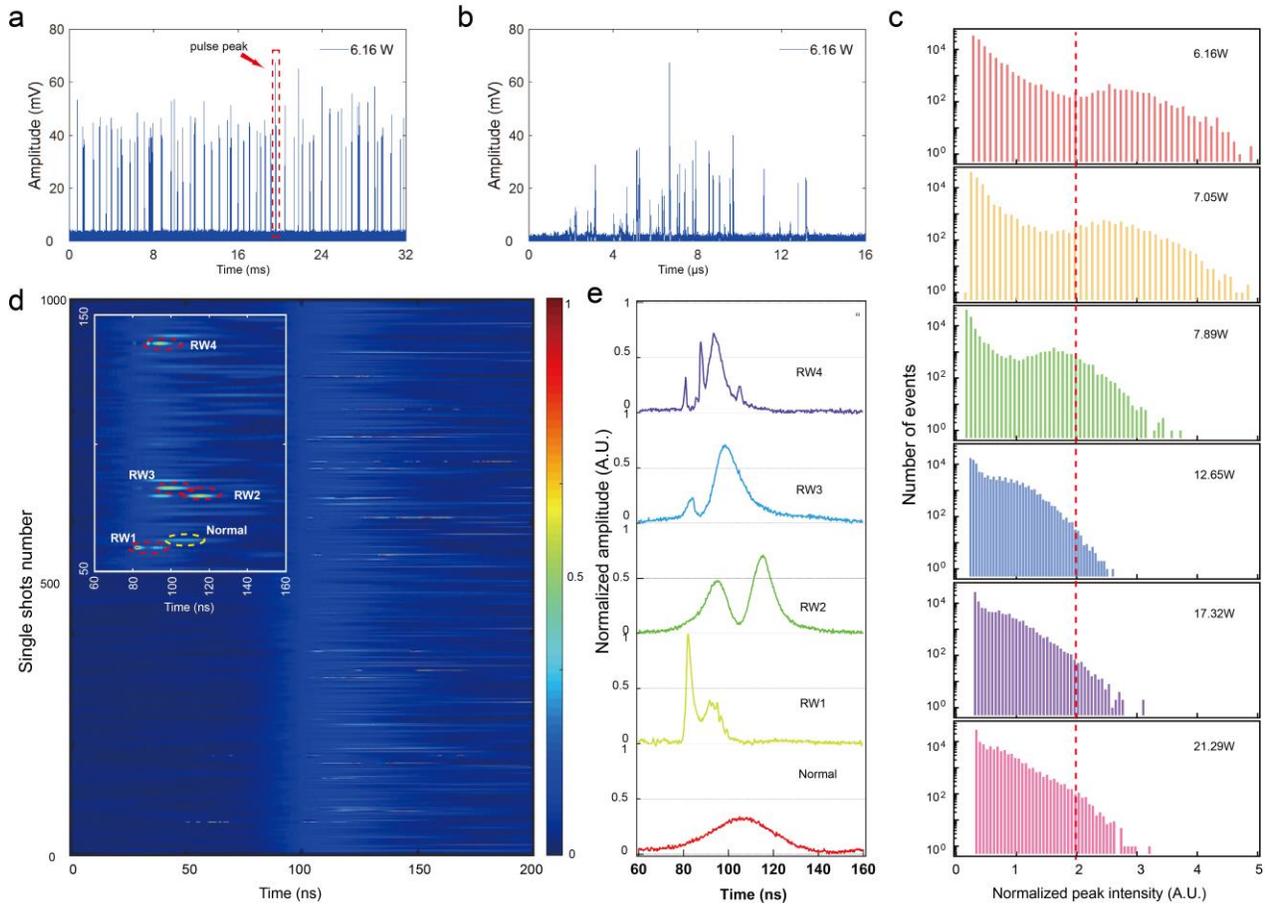

Figure 3 | Temporal characteristics and statistical features of output random laser. (a) and (b) Typical temporal trace of the output random laser around threshold. (c) Histograms (log scale) showing the distribution of optical intensity maxima for $10^5$ trace events. The peak intensities are normalized by the corresponding SWH values. The vertical dash lines indicate the 2×SWH and thus mark the limit for a pulse to be considered as optical RW. (d) Evolution of consecutive traces around optical RW events at threshold. (e) Pulse shapes of typical normal and optical RW events.

To check whether the RWs are generated or not in this self-pulsed RFL, we measured and recorded the peak amplitude of the shot-to-shot nanosecond pulse and added it to a histogram. The measurement and recording are similar with those in [34] and automatically repeated at a rate of about $2 \times 10^5$ trace samples per hour. Figure 3(c) presents the statistical distribution histograms recorded for the peak amplitudes of $10^5$ consecutive traces corresponding to different pump levels. Additionally, the vertical dotted line indicates the peak amplitude twice the significant wave height (SWH, defined as the mean height of the highest third of events [35]) of each measurement and denotes the calculated amplitude from which events are considered RWs. It should be noted that the peak amplitudes shown as horizontal axis of Fig. 3(c) are normalized by corresponding SWHs to investigate the proportion evolution of RW in the power boosting more conveniently. The heavy-tailed histograms display numerous events with peak amplitudes larger than twice the SWH, thus revealing the generation of RW behavior in this self-pulsed RFL. Furthermore, around the threshold, the highest recorded peak amplitudes can reach intensity about 5 times that of the corresponding SWH, and the emergence proportion of RW events are about 4.1 %. As the pump power increases, the RW events can always be observed even at maximal power level. However, the peak amplitude and generation probability decreases to stable values of ~3 times that of respected SWH and 0.35 %. These decrements may be induced by the denser nanosecond pulses and associated higher SWH value. To show the trend of RW events generation in this self-pulsed RFL at 6.16 W pump level, the shot pulse evolution over one thousand consecutive traces and typical enlarged view around RW events are depicted in Fig. 3(d). Firstly, on account of stochastic nature of SBS, the recorded traces suffer from shot-to-shot perturbations. Thus, there are rare high amplitude traces with peak intensities higher than twice the SWH, which are also generated unpredictably and disappeared with a transient state. Indeed, these agree well with the criteria of RW behavior [34, 36]. The typical pulse shapes of normal and RW events, illustrated by Fig. 3(e), also indicate the high fluctuation and multiplicity of pulse duration and amplitude of RW events. The respected statistical result of pulse durations of these shot pulses is given in Fig. S2 of the supplementary figures.

Contrasted with the unstable nanosecond pulsations with RW characteristics induced by intrinsic stochastic nature of SBS effect, the average powers of output lasers (measured by power meters with 1 second/ point resolution and half an hour recorded length, more details can be found in Fig. S3 of the supplementary figures) indicate good long-term power stabilities, which may be benefitted from the employment of ultrastable ASE source as pump light [37, 38]. This situation is somewhat similar to the classical physical problem in which long-term average properties of the system are determined by the random behavior of a huge number of waves [39]. In hence, this incoherently pumped self-pulsed RFL exhibits good long-term average power stabilities at quasi-hour scale and giant pulse behaviors with RW characteristics at nanosecond scale simultaneously.

**Radiofrequency evolutions of self-pulsed RFL.** To investigate the self-pulsed operation dynamics and confirm the generation of SBS effect, we measure the radio frequency (RF) evolutions of output random laser, as depicted in Fig. 4. From the lasing threshold to the maximal pump level, well-pronounced beating peak near 15 GHz, which coincides with the Brillouin frequency shift ($\Delta v_B = 2nv_a/\lambda$, where n, $v_a$ and $\lambda$ are respectively the refractive index, the velocity of acoustic wave and the pump wavelength of the SBS process) of laser around 1120 nm in the passive fiber [19, 29], can always be measured clearly and indicate the presence of SBS effect in the power boosting process. Additionally, Fig. 4(b) shows the enlarged views around the 15 GHz characteristic peak. At the lasing threshold, independent peaks can be observed with > 30 dB signal-to-noise ratio (SNR) for the highest signal. However, with the scaling of pump power to 7.05 W, the RF spectroscopy develops more components appear at new frequencies, which is similar to that in [40], and evolves to quasi-cw shape with little fluctuations. In the power boosting of pump power from 7.05 W to maximal 21.29 W, the gain magnification of the low frequency section (lower than ~15 GHz) in the RF spectroscopy is superior to that of the high frequency section in respect to the cascaded SBS process generated long wavelength components and corresponded lower RF frequency (as indicated by the formula before). At the maximum pump level, the 10 dB width of the broadband RF spectral envelope is about 285 MHz (14.725 GHz to 15.110 GHz), corresponding to the wavelength from 1142.91 nm to 1113.79 nm. This RF spectroscopy measurement coincides well with the former optical spectrum test plotted in Fig. 2(b).

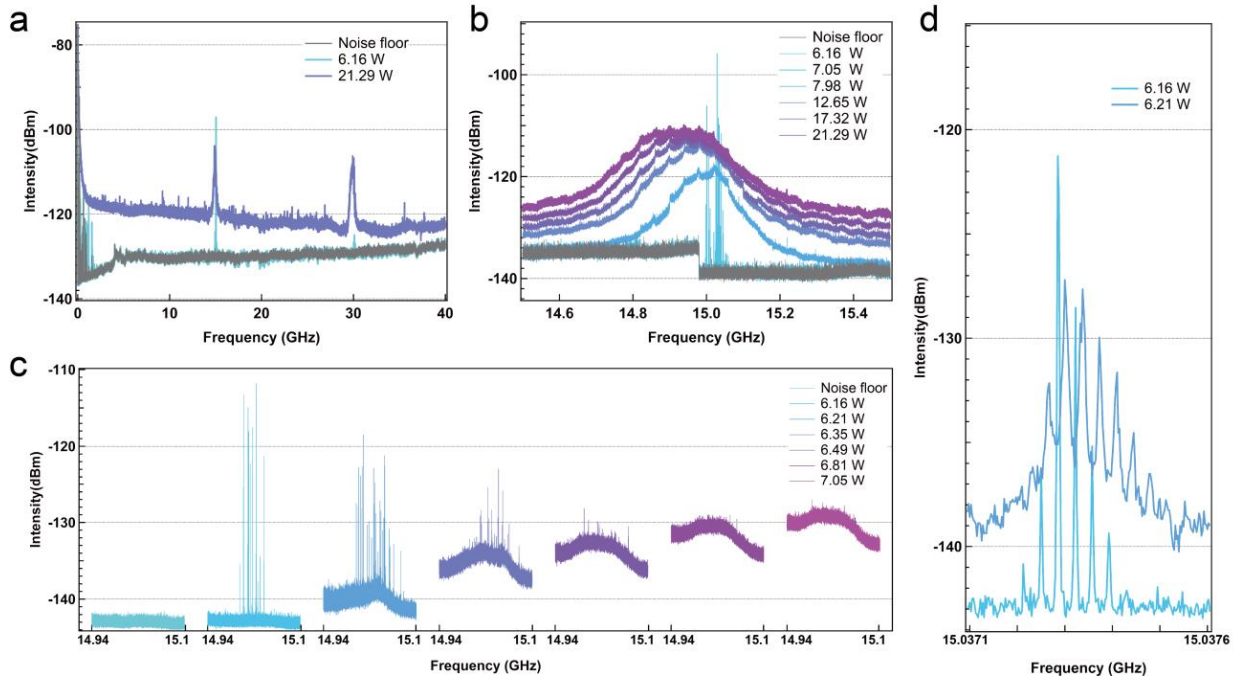

Figure 4 | RF spectra of output random laser. (a) Long span RF spctra (measured with a resolution of 1 kHz). (b) The RF spectral envelopes corresponding to SBS effect (recorded with a resolution of 10 Hz). (c) The spectra evolve from mass independent peaks to quasi-cw envelopes (measured with a resolution of 1 Hz). (d) Enlarged view of selected RF peaks around lasing threshold.

Additionally, we also investigate the RF spectroscopy evolution around the characteristic frequency corresponding to SBS effect from mass independent peaks to quasi-cw spectral envelope, as charted in Fig. 4(c) and

(d), and can find the coexistence states of these two typical RF spectra. The amplitudes of the quasi-cw spectral pedestal and the signal peak increases and decreases gradually and inversely with the enhancing of pump power from threshold to 7.05 W. Additionally and surprisingly, specific fine structure with frequency interval of about 36 kHz can be found around the lasing threshold, as displayed in Fig. 4(d). This RF spectra modulation will be discussed later. And the conversion of RF spectroscopy to quasi-cw spectral envelope may be explained as follows. Despite that only independent peaks are generated at the lasing threshold, with the pump power scaling, multiple nonlinear [26] (such as four-wave-mixing, self-phase modulation and cross-phase modulation) interactions and associated spectral broadening can generate new wavelength components as pump light of SBS process and bridge the gaps between the independent peaks. As the characteristic RF peaks near 15 GHz are relevant to the pump wavelength of SBS process in the passive fiber, indicated by the former Brillouin frequency shift formula, the gaps bridging of pump spectrum can also smooth the RF spectroscopy and induce the envelope. In hence, with the aid of cascaded SBS process and multiple nonlinear effects, the mass independent peaks at the lasing threshold can evolve to quasi-cw spectral envelope gradually with the power scalability.

**Threshold-like beating peak resolvable behavior.** Longitudinal modes in a regular laser with defined cavity structure, such as Raman fiber laser, are determined by the cavity length and can be resolved in the RF spectroscopy. In contrast, previous works on RFL predicted that the cavity-free feature of RFL can induce unique mode-beating free characteristic of output random laser [8, 28, 41]. However, we observe clear beating peaks in a limited power range near lasing threshold in the RF spectra of this presented RFL, as plotted in Fig. 5. Despite that no characteristic component can be found in the RF spectroscopy below the lasing threshold, resolvable beating peak centered at about 36 kHz, which agrees well with the modulation interval at the 15 GHz characteristic frequency of SBS effect (as depicted in Fig. 4(d)), can be identified at pump power of 6.11 W. The potential dynamics of this amazing phenomenon will be discussed later. Additionally, compared with the spectral separation of the longitudinal mode in Raman fiber laser [39], which can be estimated to about 33.68 kHz for 3.05 km passive fiber by the classical formula $c/2nL$ ($c$ is the speed of light in vacuum, $n$ is the refractive index of the fiber core and $L$ is the passive fiber length), the beating peak of this RFL has slightly difference. The characteristic frequency increment may lie on the decreasing of effective length, which is defined by $L_{eff}= [1-\exp(-\alpha L)]/\alpha$ ($\alpha$ is formed as a result of specific gain and loss profiles jointly) [28]. It should be noted that the effective length $L_{eff}$ of passive fiber cannot be calculated precisely right now for the complicated spatiotemporal distributions of transient Raman and cascaded Brillouin Stokes waves and associated transient gain and loss profiles in this self-pulsed RFL.

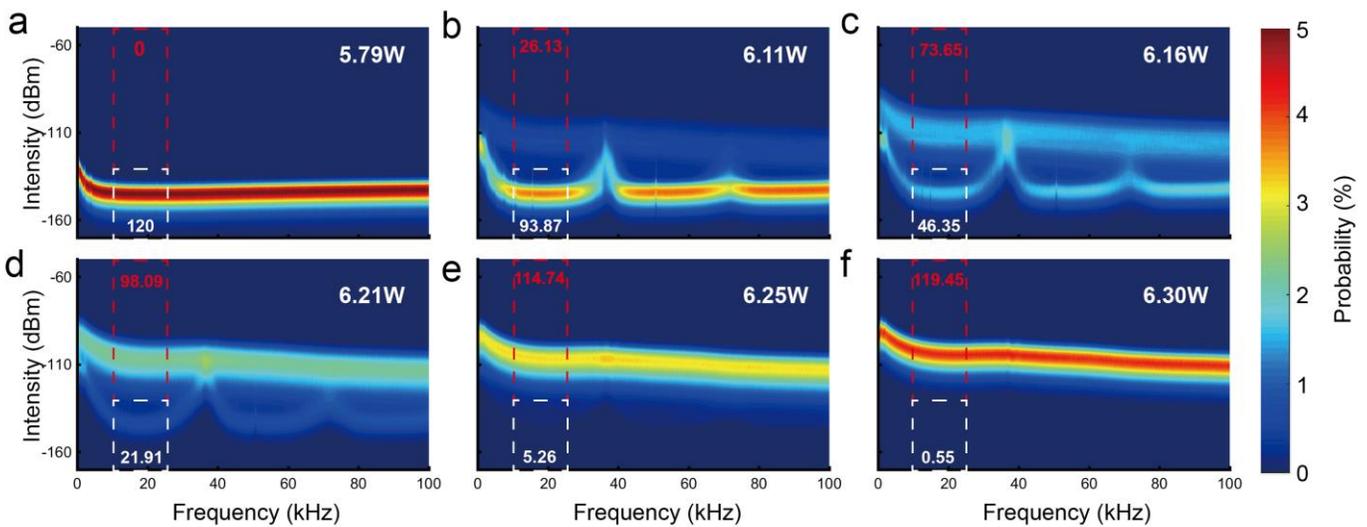

Figure 5 | Probability bitmaps of self-pulsed random laser at different pump levels around lasing threshold. The bitmaps display the signal generation density in percent at corresponding pixel with resolutions of 125 Hz/pixel and 0.6 dBm/pixel for the lateral and vertical axis, respectively. And they are calculated basing $1.22\times10^5$ measurements in ten seconds consecutive operation.

In particular, similar to [42] in a sense, an experimental indication of the coexistence of two lasing states can be observed in Fig. 5(b)- beating-peak-resolvable state and beating-peak-free state. To characterize the composition evolution between these two states quantitatively while pumping conditions are changed, we record the generation probability of the RF spectra at corresponding pump levels in 10 seconds continuous operation automatically with the aid of a real-time signal analyzers and calculate the probabilities in the defined boxes (from 10 kHz to 25 kHz). As the sum of the probabilities for each recorded frequency component with different intensities should be 100 %, and the defined box possesses 15 kHz width and corresponded 120 pixels (125 Hz/ pixel), the probability sums in the defined boxes are 120. Then, the probability sum in the white and red dashed line box approximately reveals the relative emergence intensity of beating-peak-resolvable state and beating-peak-free state, respectively. Before the lasing threshold (e. g. 5.79 W), only spontaneous Raman emission component can be measured in the output optical spectrum, as illustrated in Fig. 2(b), and no characteristic frequency can be found in the RF spectroscopy. With the boosting of pump level from 6.11 W to 6.30 W, the probability sum in the red and white box increases and decreases, inversely, indicating the transition from beating-peak-resolvable state to beating-peak-free state. At last, the beating peak could almost disappeared at 6.30 W pump level, which is similar to the longitudinal mode structure washing out phenomenon in classical Raman fiber laser in [39]. The evolution of beating-peak-free state to absolutely dominant component may be caused by the nonlinear interactions led collective mode formation [7]. At last, we would like to stress that the amazing fact here is the beating-peak-resolvable state near the lasing threshold. Further theoretical and experimental investigation is necessary to reveal the exact physical dynamics underlying the appearance and disappearance of this state for such cavity-free RFL.

**Discussion**

Note that SBS effect had been observed in many RFLs constructed by only common passive fiber in the past years [7, 33], and the previous literatures had indicated that SBS effect and associated irregular pulses and spiky optical spectrum could be suppressed with the scalability of pump power overcoming lasing threshold [7, 26, 28] (e. g. about 25 % higher than lasing threshold in [7]). The respected probable reason was attributed to the pump laser intensity fluctuations induced phase fluctuations and corresponding optical spectral broadening via the cross-phase-modulation process [7, 26, 28]. Nevertheless, in this RFL pumped by incoherent ASE source with a FWHM linewidth as broad as about 16 nm, the SBS factor can sustain from lasing threshold to even maximal pump level (21.29 W), which is about 2.5 times higher than the threshold. The most probable reason for that is the employing of ASE source with high temporal stability [37, 38] as pump source. Furthermore, apart from the identification of stochastic pulses with large pulse-to-pulse intensity fluctuations and associated spiky spectrum, the previous investigations did not reveal any RW behavior in such SBS-assisted self-Q-switched RFL platforms [7, 25, 27, 33]. Here, the temporal statistics measurement can evidence the RW characteristic of this self-pulsed RFL for the first time. And the RW events can also be observed even at maximal pump power with peak amplitudes of about 3 times of corresponding SWH value. However, we cannot simulate the self-pulsing operation regime involving cooperative RS-SRS-SBS processes in this RFL right now for the enormous computation induced by the complex interactions between different components, relative long kilometers level passive fiber and transient nanosecond level pulse duration demanded precise time step. And this can be our next goal.

On the other hand, despite that the previous investigations indicated the mode-beating-free characteristic of random laser from cavity-free structured RFL, interestingly we can observe threshold-like resolvable beating peak centered at about 36 kHz in the RF spectra of this presented RFL. As the previous investigations had indicated that the RFL can present rich kinetics around the lasing threshold [7, 43] (e.g. long living narrowband spectrum), and the localization mechanism can induce localized ordered operation state in disordered medium [44-47], this threshold-like beating-peak phenomenon may be another localization dynamic induced threshold behavior in RFL. The beating peak may owes to the modal interference between light localization induced threshold lasing modes, for which the modes in a disordered scattering system are quite like the modes of standard optical resonators, such as the Fabry–Perot cavity (more details can refer to [44-46] and references therein). The rigorous physical origins of such beating-peak-resolvable state in this RFL without fixed laser cavity is, however, still not fully interpreted and future research is required to better understand the rich and complex dynamics involved.

In conclusion, we have shown an incoherently pumped RFL with optical RW behavior induced by intrinsic stochastic nature of SBS effect under even maximal pump power (about 2.5 times higher than the lasing threshold), and the observation of threshold-like beating-peak behavior in the RF spectroscopy. This is the first presentation of optical RW characteristic in RFL, to the best of our knowledge, and the investigation may pave a new way to reveal the fundamental physical dynamics underlying the operation and performance of random laser in disordered media.

Further statistical and analytical studies are necessary to characterize the remarkably challenging complex phenomena in this RFL involving cascaded nonlinear interactions such as RS-SRS-SBS cooperative processes.

## Methods

**Experimental setup.** In brief, the RFL primarily composes of a piece of 3.05 km long passive fiber (G.652, 8.2 μm core diameter, and 0.14 numerical aperture) with a total transmission loss of about 3 dB for pump light that is fused with one wavelength division multiplexer (WDM). The end-facets of passive fiber and pigtailed fiber of WDM are cleaved with angles of 8° and reflectivities of about $10^{-6}$ level to suppress the Fresnel reflections and ensure that the feedback is only provided by the distributed Rayleigh scattering in the passive fiber. The pump light we employed is incoherently ASE source centered at 1063 nm with a FWHM linewidth of about 16 nm and unique little temporal fluctuation feature (more details can be referred to Ref. [38]).

**Funding**

The work is supported by the National Natural Science Foundation of China (61322505 and 61635005), Huo Ying Dong Education Foundation of China (151062), Natural Science Foundation of Hunan Province, China (2018JJ03588) and Hunan Provincial Innovation Foundation for Postgraduate Student (CX2017B030).

# SUPPLEMENTARY FIGURES

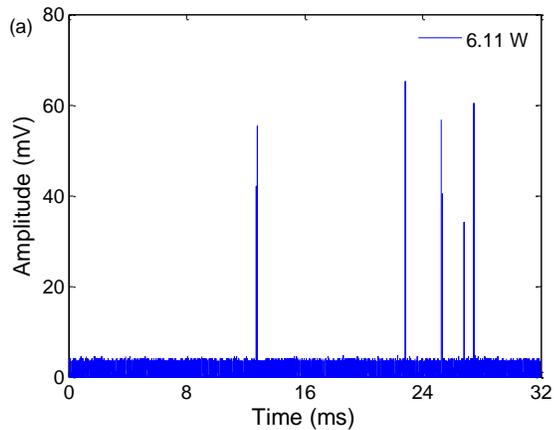

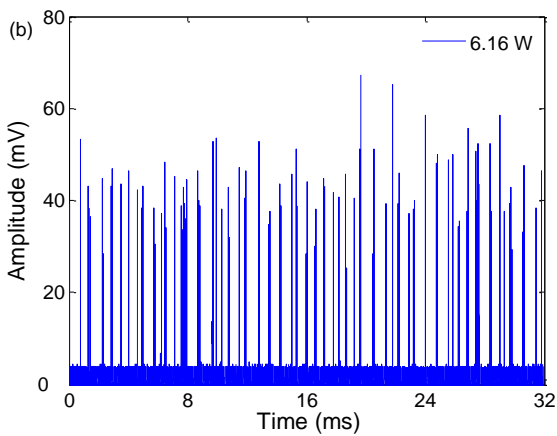

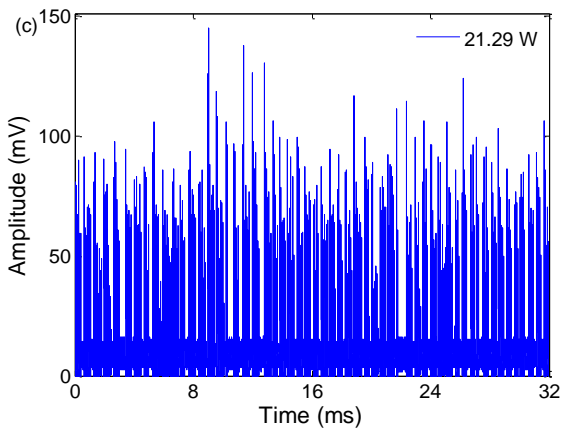

Supplementary Figure 1 | Typical temporal traces of the output random lasers at three different pump levels. With the boosting of pump power from lasing threshold to maximal level, the pulse peak amplitude increases, and the stochastic pulses become denser.

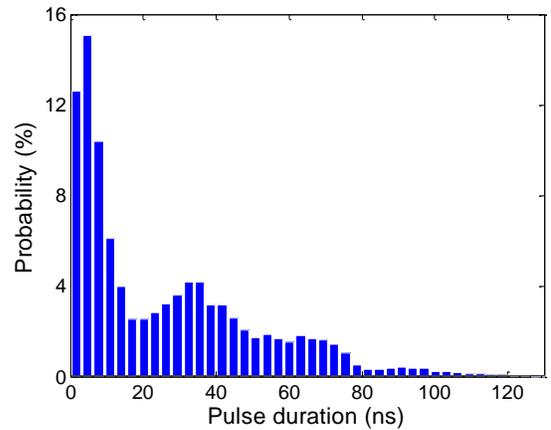

Supplementary Figure 2 | Distribution of pulse duration. This histogram, built on one thousand consecutive trace measurements, presents the statistics of the pulse duration.

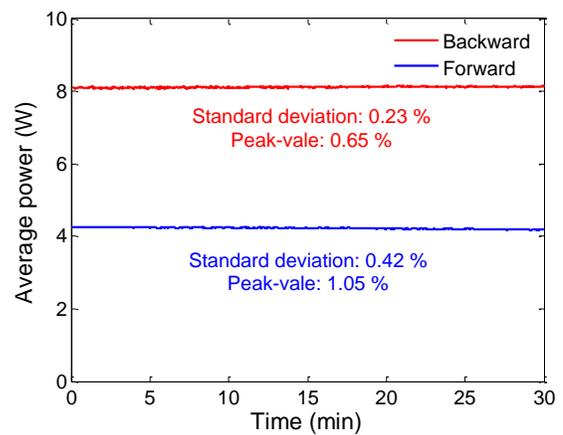

Supplementary Figure 3 | Long-term average power stabilities of output lasers. The average powers are measured by power meters with 1 second/ point resolution and half an hour recorded length.